\documentclass[12pt]{article}
\usepackage{graphics}
\begin{document}
\begin{center}
{The Non-perturbative term for the Vector Form Factor of Pion Decay}
\end{center}
\begin{center}
{Susumu Kinpara}
\end{center}
\begin{center}
{\it Institute for Quantum Medical Science (QST)\\ Chiba 263-8555, Japan}
\end{center}
\begin{abstract}
The vector form factor of the decay $\pi^{+} \rightarrow \gamma + e^{+} + \nu_e$ is calculated 
by the method for the pseudovector pion-nucleon system.
The non-perturbative term is taken into account by using the parameter for the self-energy of nucleon
following our previous calculation of the pion form factor.
The suppression of the anomalous interaction of proton in the loop integral is significant to understand the experimental value.
\end{abstract}
\section*{\normalsize{1 \quad Introduction}}
\hspace*{4.mm}
Many phenomena related to nucleon are influenced by pion which plays the decisive role to mediate the nuclear force particularly in the intermediate energy regions.
In addition to the process of the exchange of pions it is plausible that the nucleon does not exist singly 
and surrounded by pions to interact with other bosons appropriately.
These experimental facts are represented by the structure constants and the method of the field theoretical calculation is required to explain
the respective values quantitatively. 
\\\hspace{4.mm}
It has been known that there are two types of the interaction for the pion-nucleon system that is the pseudoscalar and the pseudovector couplings.
Although the latter is used to proceed the calculation of the process these interactions are connected with each other by the self-energy of nucleon.
Then the correction of it to the vertex is essential to examine the cause of the gap from the experiment.
The non-perturbative term is ascribed to the interaction of the pseudovector coupling.
The non-perturbative relation under the equation of motion makes us possible to derive it.
It is independent of the cancellation between the non-covariant interactions in the procedure of the perturbative expansion for the pseudovector coupling lagrangian.
\\\hspace{4.mm}
In the framework of the meson-exchange model the calculation of the higher-order diagram is connected with the self-energy 
and it is interesting to investigate the effect on the structure constant of each process.
The anomalous magnetic moment of nucleon is typically one of the constants and it is obtained by the use of the identity of the non-perturbative relation 
for the on-shell states of the nucleons.
The result of the calculation yields the numerical value which is in agreement with the experimental value assuming the decrease of the coupling constant.
The correction of the pion propagator with the polarization function would supply the factor appropriate to the region of the pion momentum.  
\\\hspace{4.mm}
When the nucleon is at the off-shell state the photon-nucleon-nucleon vertex does not have the usual form of the anomalous interaction.
The change of the vertex is probably stemmed from the higher-order diagrams of the self-energy.
In order to take account of the correlations which are unknown at present
the parameters of the phase shift in the pion-nucleon elastic scattering at the low-energy regions have been used to determine the self-energy.
The numerical value is not large in comparison with that of the perturbative treatment and then it is favorable for the correction
to the loop diagram which is based on the vertex of the pseudoscalar coupling. 
\\\hspace{4.mm}
The inclusion of the self-energy in the nucleon loop is expected to give the value of the structure constant good enough for us.
In actual the photon-pion-pion three-point vertex in the electron-pion elastic scattering is roughly understood by the pseudoscalar model
and the non-perturbative term plays a role of the correction to the lowest-order calculation \cite{Kinpara}.
In the present study the effect of the modification of the vertex is investigated applying to the radiative decay of pion.
\\
\section*{\normalsize{2 \quad The calculation of the vector form factor}}
\hspace{4.mm}
The process $\pi^{+} \rightarrow \gamma + e^{+} + \nu_e$ is expressed by the vector and the axial-vector form factors $F_V$ and $F_A$ \cite{Bryman}
associated with the vertices to give the parts of the vector and the axial-vector currents
in the diagram of the nucleon loop to construct the invariant amplitude.
In the case of $F_V$ the loop contains the gamma matrix $\gamma_5$ and the form of the amplitude is determined by using the W-T identity
which connects the photon-nucleon-nucleon vertex $\Gamma(p_f,p_i)$ between the initial and the final momenta $p_i$ and $p_f$ to the nucleon propagator $G(p)$ as 
\begin{eqnarray}
(p_f-p_i) \cdot \Gamma(p_f,p_i) = \tau_p ( G(p_f)^{-1} - G(p_i)^{-1} )
\end{eqnarray}
where $\tau_p \equiv (1+\tau_3)/2$ distinguishes proton and neutron in the isospin space and the right side of Eq. (1) acts on the proton.
The part of the loop $\Pi_{\mu\rho}$ suffices the relation $q^\mu \, \Pi_{\mu\rho} = 0$ with $p^\prime = p - q$
in which $p$ and $q$ are the momenta of the pion and the emitted photon. 
The simple form is attributed to the property of the trace which consists of four gamma matrices besides $\gamma_5$ 
and only the form $\Pi_{\mu\rho} \sim \epsilon_{\mu\rho\nu\sigma} q^\nu p^\sigma$ is possible.
\\\hspace{4.mm}
The value of $F_V$ depends on the model which is used for the calculation of the loop diagram.
The common feature to the decay of the neutral pion to two photons indicates 
that the approximation of the photon-nucleon-nucleon vertex $\Gamma_\mu \rightarrow \gamma_\mu$ is not sufficient to understand the decay process quantitatively.
A way to include the effect of
the right side in the W-T identity is to use the anomalous interaction $\sim \sigma_{\mu\nu}q^\nu$.
The coefficient gives the magnetic moment of nucleon and the interaction is assumed to be exact when the on-shell condition is imposed on both sides of the vertex.
Then the vertex of the electromagnetic interaction is approximated to $\Gamma_\mu \approx \gamma_\mu - \kappa i \sigma_{\mu\nu}q^\nu /2 M$
with $\sigma_{\mu\nu} = i [\gamma_\mu,\gamma_\nu]/2$ and a constant $\kappa$.
\\\hspace{4.mm}
The pion-nucleon-nucleon vertex is corrected by the perturbative calculations 
and it has been shown that the result of the higher-order does not give the large effect as long as the on-shell condition is used to the nucleon.
Another possible extension of the vertex is the non-perturbative term about the pseudovector coupling interaction.
The vertex part is modified as
\begin{eqnarray}
\gamma_5 \gamma \cdot p \rightarrow \gamma_5 \gamma \cdot p + G(p+k)^{-1} \gamma_5 + \gamma_5 G(k)^{-1} 
\\\nonumber\\
= -2 M \gamma_5 - \Sigma(p+k) \gamma_5 -\gamma_5 \Sigma(k)
\end{eqnarray}
using the self-energy of the nucleon propagator $G(p) = (\gamma\cdot p -M -\Sigma (p) )^{-1}$.
It is ascribed to the non-perturbative relation arising from the equation of motion for the pion-nucleon interacting system 
and different from the non-covariant terms for the perturbative calculation which are cancelled with each other.
The effect of the second and the third terms in Eq. (3) is examined in the next section. 
\\\hspace{4.mm}
The loop integral $\hat{T}_{\mu\rho}(p^\prime,p)$ for the vector part of the weak interaction is related to the vector form factor $F_V$ as 
\begin{eqnarray}
- i \sqrt{2} m \, G \, \hat{T}_{\mu\rho}(p^\prime,p) = F_V \, \epsilon_{\mu\rho\nu\sigma} \, q^\nu p^\sigma
\end{eqnarray}
\begin{eqnarray}
\hat{T}_{\mu\rho}(p^\prime,p) \equiv \int \frac{d^4 k}{i (2 \pi)^4} \frac{T_{\mu\rho}(k,p^\prime,p)}{(k^2-M^2)((p^\prime+k)^2-M^2)((p+k)^2-M^2)}
\end{eqnarray}
\begin{eqnarray}
T_{\mu\rho}(k,p^\prime,p)\equiv {\rm Tr}[(\gamma\!\cdot\! k+M) \gamma_\rho (\gamma\!\cdot\! (p^\prime+k)+M) \Gamma_\mu (\gamma\!\cdot\! (p+k)+M) \gamma_5 ]
\end{eqnarray}
The strength of the pion-nucleon-nucleon vertex is $G \equiv 2 M f / m$ with the pseudovector coupling constant $f=1$, the pion and the nucleon masses $m$ and $M$.
The sign of the tensor $\epsilon_{\mu\rho\nu\sigma}$ is chosen as $\epsilon_{0123}=1$   
and $\gamma_5$ is defined by the standard one $\gamma_5 = i \gamma^0 \gamma^1 \gamma^2 \gamma^3$ in the present study. 
\\\hspace{4.mm}
To prepare the numerator $T_{\mu\rho}(k,p^\prime,p)$ in Eq. (5) the terms linearly dependent on $k$ and does not contain $M$ factor arising from the anomalous interaction are dropped 
since they result in the higher-order of the $F_V$ taking the form of $\sim p p/M^2$ or $\sim p q/M^2$ ($p \sim q \sim m$)
and then smaller than the main parts.
For the present it is sufficient to evaluate the trace including four and six $\gamma$ matrices along with a $\gamma_5$ in it.
The calculation of the trace is performed 
by using the identity $\gamma_\mu \gamma_\nu \gamma_5 = g_{\mu\nu} \gamma_5 - \epsilon_{\mu\nu\rho\tau} \sigma^{\rho\tau}/2$ 
to lower the number of the $\gamma$ matrices and eliminate $\gamma_5$. 
Taking the above consideration into account $T_{\mu\rho}(k,p^\prime,p)$ is given as follows
\begin{eqnarray}
&&T_{\mu\rho}(k,p^\prime,p) = - 4 \, i \, M \, \epsilon_{\mu\rho\nu\sigma} \, q^\nu p^\sigma 
- 2 \, i \, M \, \kappa \, \epsilon_{\mu\rho\nu\sigma} \, q^\nu (P+k)^\sigma   \nonumber\\\nonumber\\ 
&&- \frac{i \kappa}{M} q^\nu [\, 2 \epsilon_{\mu\rho\tau\sigma} {p^\prime}^\tau k_\nu k^\sigma + \epsilon_{\nu\tau\mu\sigma}p^\tau k_\rho k^\sigma 
\nonumber\\\nonumber\\ 
&&\qquad\qquad+\epsilon_{\mu\rho\tau\nu}(p^\prime \cdot k k^\tau + p^\tau k^2 + k^\tau k^2 ) -(\mu\leftrightarrow\nu)]
\end{eqnarray}
where $\kappa$ denotes the strength of the anomalous interaction and $P \equiv p + p^\prime$.
\\\hspace{4.mm}
Regarding the $k$-integral the following relations are used to evaluate the terms $k_{\mu\nu}$ and $k_{\mu\nu\rho}$
defined by $k_{\mu\nu\rho \, \cdots} \equiv k_\mu k_\nu k_\rho \cdots$ in $T_{\mu\nu}$
\begin{eqnarray}
\hat{k}_{\mu\nu} = -\frac{g_{\mu\nu}}{8 (4 \pi)^2} \, (1-2 \, {\rm log}\frac{{\it\Lambda}^2}{M^2})
\end{eqnarray}
\begin{eqnarray}
\hat{k}_{\mu\nu\rho} = \frac{1}{36 (4 \pi)^2} \, (P_\mu g_{\nu\rho} +P_\nu g_{\rho\mu} +P_\rho g_{\mu\nu}) \, (1-3 \, {\rm log}\frac{{\it\Lambda}^2}{M^2})
\end{eqnarray}
giving rise to the divergence.
In Eqs. (8) and (9) the terms of the higher-order $O((\frac{p}{M})^2)$ are neglected.
The definition of the $k$-integral $\hat{k}_{\mu\nu}$ and $\hat{k}_{\mu\nu\rho}$ is similar to that of Eq. (5).
When they are divergent it is necessary to introduce the cut-off $\it\Lambda$ to make them convergent.
Each divergent integral has the respective cut-off parameter different from each other because of the finite value unsettled.
To determine it and connect them the cut-off $\it\Lambda$ has been taken from the integral in our previous study \cite{Kinpara}.
It enables to compare the divergent quantities of the integrals.
\\\hspace{4.mm}
The calculation of $F_V$ results in 
\begin{eqnarray}
F_V = \frac{4 \sqrt{2} f}{(4 \pi)^2} \,[\, 1+\frac{1}{4}(\kappa_p+\kappa_n)(1+2 \, {\log}\frac{{\it\Lambda}^2}{M^2}) \,]
\end{eqnarray}
neglecting the order of $\sim m^2/M^2$.
The cut-off $\it\Lambda$ is expected to be ${\it\Lambda} \sim M$, so that it would give only the minor effect. 
The order of the integrand $k \sim -3$ makes us understand the appearance of the divergence different from the calculation of the pion charge radius
in which the $\it\Lambda$ appears only in the symmetric part under the exchange of the subscripts $\mu \leftrightarrow \nu$ \cite{Kinpara}.
In the case of $F_V$ the anomalous interaction gives the dependence $\sim \kappa_p + \kappa_n$.
Then the most part is cancelled with supposing the coefficients of the magnetic moments ($\kappa_p$ = 1.79, $\kappa_n$ = $-$1.91).
The sign in front of $+\,\kappa_n$ is definite in view of the connection between the two loops of the different directions
which are converted mutually by the transformation $p \rightarrow -p$ and $p^\prime \rightarrow -p^\prime$.
\\
\section*{\normalsize{3 \quad The effect of the non-perturbative term }}
\hspace{4.mm}
For the calculation of the loop integral an advantage of the self-energy in the non-perturbative term 
is that the effect is interpreted as the correction to the result of the pseudoscalar interaction.
It has been verified that the pseudoscalar model is derived from the pseudovector coupling pion-nucleon interaction 
and applied to the calculation of the form factor of pion.
The pion-nucleon-nucleon vertex is modified as 
\begin{eqnarray}
\gamma_5 \rightarrow (1+c) \gamma_5 + \frac{c}{2 M} \gamma_5 \gamma\cdot (a-b)
\end{eqnarray}
where $a$ and $b$ are the outgoing and the incoming momenta of nucleons respectively.
In the case of the decay of pion with the momentum $p$ the difference is $a-b = p$ and does not raise the order of the integrand $k \sim -3$.
\\\hspace{4.mm}
The self-energy $\Sigma(a) = M c_1(a) - \gamma\cdot a \, c_2(a)$ is substituted to the approximated form
as $c_1(a) \approx c_2(a) \approx c$ neglecting the higher-order terms of the expansion in $a^2 - M^2$. 
When the variable of the integral $k$ takes the value $k^2 \sim M^2$ the approximation is valid and enables to choose the value $c$
which characterizes the process of the decay.
The idea is supported by the value of the cut-off $\it\Lambda$ which is expected to take $\it\Lambda \sim M$ 
so as to include possibly the shape of the exact propagator of nucleon in the loop diagram.
\\\hspace{4.mm}
Following the extension of the vertex in Eq. (11) the numerator of the loop integral $T_{\mu\rho}(k,p^\prime,p)$ is modified as 
\begin{eqnarray}
T_{\mu\rho}(k,p^\prime,p) \rightarrow (1+c) \, T_{\mu\rho}(k,p^\prime,p) + \frac{c}{2} \, T^{\,\prime}_{\mu\rho}(k,p^\prime,p)
\end{eqnarray}
\begin{eqnarray}
&&T^{\,\prime}_{\mu\rho}(k,p^\prime,p) = 4 \, i \, M^{-1} \, \epsilon_{\mu\rho\nu\sigma} \, [\, -M^2 \, {p^\prime}^\nu p^\sigma +M^2 \, k^\nu p^\sigma
-2 k \cdot p \, k^\nu {p^\prime}^\sigma \nonumber\\\nonumber\\
&&-k^2 \, (p^\prime+k)^\nu p^\sigma \, -p^2 k^\nu {p^\prime}^\sigma \,] 
+2 \, i \, \kappa \, M^{-1} \, q^\nu [ \, (M^2+k^2) \epsilon_{\mu\rho\nu\sigma} p^\sigma \nonumber\\\nonumber\\ 
&&+2 p^\tau k^\sigma( \epsilon_{\mu\sigma\nu\tau} k_\rho 
- \, \epsilon_{\nu\rho\tau\sigma} k_\mu + \, \epsilon_{\mu\sigma\nu\rho} k_\tau + \, \epsilon_{\mu\rho\tau\sigma} k_\nu) \,]
\end{eqnarray}
where the linear term in $k$ is dropped as well as the $T_{\mu\rho}(k,p^\prime,p)$.
Performing the $k$-integral the $F_V$ is modified as
\begin{eqnarray}
F_V \rightarrow F^c_V \equiv (1+c) \, F_V - \frac{3 \sqrt{2} f}{(4 \pi)^2} \,[\,1+\frac{1}{3}(\kappa_p+\kappa_n) \,] \, c 
\end{eqnarray}
including the effect of the non-perturbative term.
It has been verified that the divergent terms cancel out and the cut-off $\it\Lambda$ does not appear in the second term.
\\\hspace{4.mm}
It is interesting to examine the numerical value of $F_V$ and compare with the experimental value.
When $\it\Lambda$ = $M$, it yields $F_V$ = 0.035 twice as large as the experimental value $F_V^{\rm{exp}}$ = 0.017$\pm$0.008 \cite{PDG}.
The calculation gives the value large excessively similar to the case of the radius of pion.
The feature of the results would be improved by the additional terms to correct the loop diagrams.
Inclusion of the coefficient $c$ which is generated by the non-perturbative term is an effective way 
and the use of the phenomenological value $c$ = $-$0.39 changes to smaller value $F_V$ = 0.031.
\\\hspace{4.mm}
The numerical value of $c$ has been determined by the low-energy parameters of the phase-shift analysis for the elastic pion-nucleon scattering.
The off-shell behavior of the vertex is taken into account  
and therefore thought to be realistic in comparison with the use of the vertex which consists of the on-shell nucleons. 
Because of the cut-off $\it\Lambda \sim M$ the variable of the integral $k$ seems likely to exist around $k^2 \sim M^2$ 
and only the lowest-order of the expansion in $\gamma\cdot p-M$ remains to construct the self-energy $\Sigma(p)$ with a momentum $p$.
Then $\Sigma(p) = M c_1(p^2) - \gamma\cdot p \, c_2(p^2)$ is replaced by
\begin{eqnarray}
\Sigma(p) = \frac{1}{2 M}(c_2^{(0)}-4 M^2 c_2^{(1)}+4 M^4 (c_1^{(2)}-c_2^{(2)})) (\gamma\cdot p -M)^2 
\end{eqnarray}
neglecting the terms of the higher-order $O((\gamma\cdot p -M)^3)$.
The $c_{1}^{(n)}$ and $c_{2}^{(n)}$ are the coefficients of the series expansion in $p^2-M^2$ for $c_{1}(p^2)$ and $c_{2}(p^2)$.
\\\hspace{4.mm}
For proton the anomalous part of the magnetic moment is divided into three parts 
that is the part of the self-energy ($\kappa_s$), the pion current and the nucleon current.
The change of the first part makes us possible to use the other values of the coefficients in the anomalous interaction.
The value of $\kappa_s$ is approximated to
\begin{eqnarray}
\kappa_s = c_2^{(0)}-4 M^2 c_2^{(1)}+4 M^4 (c_1^{(2)}-c_2^{(2)}) \rightarrow c_2^{(0)} (= c)
\end{eqnarray}
with respect to the self-energy in Eq. (15).
\\\hspace{4.mm}
The $n=2$ model is suitable for the magnetic moment of nucleon as has been shown in our previous study.
Using the set of the parameters the relation $\kappa_s = c$ is satisfied exactly since the parameters are zero except $c \equiv c_2^{(0)}$ and $c_1^{(1)}$.
Regarding the set determined by the method of the matrix inversion to reproduce the phase-shift, 
the parameters of the higher-order remain in Eq. (16).
These parameters are eliminated from it by the above consideration on $k^2$.
Consequently the strength of the anomalous interaction of proton is changed from $\kappa_p$ to $\kappa_p^\prime$ $\equiv$ $\kappa_p -2 + c$.
On the other hand $\kappa_n$ is intact because neutron does not have the part $\kappa_s$.
\\\hspace{4.mm}
Applying the above value of $\kappa_p^\prime$ the $F_V$ results in $F_V$ = 0.010.
The effect of $\kappa_p^\prime$ reduces the value $F_V$ excessively and required to find out some other effects to recover from the underestimate.
By moving the cut-off $\it\Lambda$ from the expected value $\it\Lambda$ = $M$ to the lower direction about 12$\%$ 
it is attainable to reproduce the value of $F_V^{\rm{exp}}$.
The procedure of the loop integral by means of the use of the parameter $c$ is effective also 
against the result of the calculation for the charge radius of pion $r_\pi$ = 1.06 fm 
with $c$ = 0, that is, the use of the pseudoscalar interaction. 
It is larger than the experimental value given as $r_\pi^{\rm{exp}}$ = 0.672$\pm$0.008 fm \cite{PDG}.
The inclusion of $c$ = $-$0.39 reduces the value of $r_\pi$ to $r_\pi$ = 0.95 fm 
and in addition to it the substitution of $\kappa_p$ to $\kappa_p^\prime$ gives a rather smaller value $r_\pi$ = 0.567 fm than $r_\pi^{\rm{exp}}$.
In order to reproduce $r_\pi^{\rm{exp}}$ the value of $\it\Lambda$ is required to decrease about 40$\%$ as well as the case of $F_V$. 
\\\hspace{4.mm}
It is interesting to determine two parameters $c$ and $\it\Lambda$ by using Eq. (14) and the relation for $r_\pi$ in Ref.$\,$\cite{Kinpara} 
so they give the experimental values of $F_V^{\rm{exp}}$ and $r_\pi^{\rm{exp}}$ simultaneously.
The solution has not been found out at $c \, > 0$.
The result of the calculation contradicts the case of the magnetic moment of nucleon in which $c  = \,$2 is derived by the calculation for the self-energy of nucleon.
It means that the loop diagram needs the exact form of the propagator or the substitution to the parameter of the cut-off
along with the off-shell behavior of the vertex. 
At the region $c \, < 0$ the above method gives the solution ($c$,$\,\it\Lambda$) = ($-$0.13,$\,$0.97$M$).
The decrease of $\kappa_s$ through $c$ is important to understand these form factors quantitatively.
\\\hspace{4.mm}
According to the conserved vector current hypothesis the part of the vector current contains the term of the weak magnetism 
as well as the anomalous term of the electromagnetic interaction. 
The strength of the interaction $\kappa_w$ is determined by $\kappa_w$ $\equiv$ $\kappa_p - \kappa_n$ = 3.7 definitely.
It is an interesting subject to examine the effect of the weak magnetism on the vector form factor.
The modification of the numerator of the $k$-integral is 
\begin{eqnarray}
T^{w}_{\mu\rho}(k,p^\prime,p) \rightarrow (1+c) \, T^w_{\mu\rho}(k,p^\prime,p) + \frac{c}{2} \, T^{w \,\prime}_{\mu\rho}(k,p^\prime,p)
\end{eqnarray}
\begin{eqnarray}
T^{w}_{\mu\rho}(k,p^\prime,p) = 2 \, i \, \kappa_w \, M^{-1} \,[ \, \epsilon_{\mu\rho\tau\sigma} \, (k^2-M^2) \, (k-q)^\sigma \nonumber\\\nonumber\\
+ \, 2 \, \epsilon_{\nu\rho\tau\sigma}k_\mu k^\nu p^\sigma +\, 2 \, \epsilon_{\mu\rho\tau\sigma} k^\sigma k \cdot p \,]\, {p^\prime}^\tau
\end{eqnarray}
\begin{eqnarray}
T^{w \,\prime}_{\mu\rho}(k,p^\prime,p) 
= 2 \, i \, \kappa_w \,M^{-1}\,[ \, (k^2 - M^2) \epsilon_{\mu\rho\tau\sigma} - 4 \, k_\mu k^\nu \epsilon_{\nu\rho\tau\sigma} \,]\,
{p^\prime}^\tau p^\sigma
\end{eqnarray}
by the inclusion of the non-perturbative term. 
The terms which give only small values are neglected as well as $T_{\mu\rho}(k,p^\prime,p)$.
Using Eqs. (17)$\sim$(19) the vector form factor $F_V^w$ results in
\begin{eqnarray}
F_V^w = \frac{2 \sqrt{2} f}{(4 \pi)^2} \, \kappa_w \,[\,(1+c) \, (1+{\rm log} \frac{{\it\Lambda}^2}{M^2})\, - \frac{1}{2} \, c \,]
\end{eqnarray}
It is necessary to change the strength $\kappa_w$ to $\kappa_w^\prime \equiv \kappa^\prime_p - \kappa_n$ 
due to the substitution of the parameter from $c = 2$ to $-0.39$ in the loop diagram.   
\\\hspace{4.mm}
The joint use of $F_V^c$ and $F_V^w$ yields $F_V^c + F_V^w$ = 0.0288 for which the change of the anomalous interaction
from $\kappa_p$ to $\kappa^\prime_p$ with $c = -0.39$ is applied.
The above value of $F_V^c + F_V^w$ is close to the value $F_V$ = 0.0263 predicted by the measurement of the decay of the neutral pion 
under the conserved vector current hypothesis \cite{Bryman}. 
It indicates that the point interaction ($\sim \gamma_\mu$) carries the large part of the contribution 
and the higher-order pion processes represented
by the anomalous interactions are relatively small or cancelled out each other. 
The terms of $\sim \kappa_p^\prime \kappa_w^\prime$ and $\sim \kappa_n \kappa_w^\prime$ are not included in the present study.
They could shift the final result of the calculation to the favorable direction.
\\
\section*{\normalsize{4 \quad Summary and remarks }}
\hspace*{4.mm}
In the loop diagram the anomalous interaction is essential to obtain the result of the charge radius of pion by virtue of the identity on the photon-pion-pion vertex.
Also, for the pion form factor the calculation of the lowest-order has revealed that the strength of the interaction is too strong to describe it
with the values of the magnetic moments of nucleon.
An idea is to decrease the strength of the parameter in the self-energy part using the fit to the phase-shift of the pion-nucleon elastic scattering.
The anomalous interaction brings the cut-off which represents the real shape of the nucleon propagator.
The method of the conserved vector current hypothesis is useful to compare the vector form factor with the decay of the neutral pion.
The breaking of the relation implies that the model composed of pion and nucleon under the point interaction needs the further progress.
\\
\hspace{4.mm}

\end{document}